\documentclass[12pt,preprint]{aastex}

\newcommand{\figref}[1]{Figure \ref{#1}}
\newcommand{\eqref}[1]{eq. \ref{#1}}
\newcommand{\Alfven}{Alfv\'{e}n }
\newcommand{\Chandra}{\textit{Chandra} }
\newcommand{\wpe}{\omega_{pe}}
\newcommand{\wce}{\Omega_{ce}}

\newcommand{\wci}{\Omega_{ci}}
\newcommand{\V}[1]{\mbox{\boldmath $#1$}}
\newcommand{\rmax}{\rm{max}}
\newcommand{\rmin}{\rm{min}}
\newcommand{\foot}{f}
\newcommand{\overshoot}{max}
\newcommand{\Vsh}{V_{sh}}
\newcommand{\Emax}{E_{max}}
\newcommand{\Emin}{E_{min}}
\newcommand{\Eref}{E_{ref}}
\newcommand{\Pesc}{P_{esc}}
\newcommand{\Tref}{T_{ref}}
\newcommand{\phiht}{{\tilde \phi}}
\newcommand{\kev}{\rm{keV}}
\newcommand{\kms}{\rm{km/s}}
\newcommand{\pc}{\rm{pc}}
\newcommand{\mug}{\mu \rm{G}}

\shorttitle{Electron Injection at Quasi-Perpendicular Shocks}
\shortauthors{Amano \& Hoshino}

\begin{document}
\title{Electron Injection at High Mach Number Quasi-Perpendicular Shocks
: Surfing and Drift Acceleration} \author{T.Amano and M.Hoshino}
\affil{Department of Earth and Planetary Science, University of Tokyo,
7-3-1 Hongo, Bunkyo-ku, Tokyo 113-0033, Japan}
\email{amano@eps.s.u-tokyo.ac.jp}

\begin{abstract}
Electron injection process at high Mach number collisionless
quasi-perpendicular shock waves is investigated by means of
one-dimensional electromagnetic particle-in-cell simulations. We find
that energetic electrons are generated through the following two steps:
(1) electrons are accelerated nearly perpendicular to the local magnetic
field by shock surfing acceleration at the leading edge of the shock
transition region. (2) the preaccelerated electrons are further
accelerated by shock drift acceleration. As a result, energetic
electrons are preferentially reflected back to the upstream. Shock
surfing acceleration provides sufficient energy required for the
reflection. Therefore, it is important not only for the energization
process by itself, but also for triggering the secondary acceleration
process. We also present a theoretical model of the two-step
acceleration mechanism based on the simulation results, which can
predict the injection efficiency for subsequent diffusive shock
acceleration process. We show that the injection efficiency obtained by
the present model agrees well with the value obtained by \Chandra X-ray
observations of SN 1006. At typical supernova remnant shocks, energetic
electrons injected by the present mechanism can self-generate upstream
\Alfven waves, which scatter the energetic electrons themselves.
\end{abstract}

\keywords{acceleration of particles --- cosmic rays --- plasmas ---
shock waves}

\section{INTRODUCTION}
The origin of nonthermal emission observed from a variety of
astrophysical objects is still a major unresolved issue of plasma
astrophysics. These include supernova shocks, extragalactic radio
sources and active galactic nuclei. Among them, shocks of supernova
remnants (SNRs) are believed to be the most probable acceleration site
of Galactic cosmic rays. Much theoretical and observational work has
been devoted to particle acceleration processes around shock waves. One
of the most widely applied theory is diffusive shock acceleration (DSA)
theory
\citep[e.g.][]{1978MNRAS.182..147B,1978MNRAS.182..443B,1978ApJ...221L..29B}. It
has been very successful in providing a natural explanation for the
power law distributions of high energy particles observed in many
astrophysical sources. This process utilizes \Alfven wave turbulence as
the particle scatterers. Under the assumption of elastic collision
between particles and waves, energetic particles scattering back and
forth across the shock front gain a net momentum because of the
converging velocity fields. The theory was extended to include the
finite shock size and the self-consistent wave excitation and applied to
the Earth's bow shock in order to account for the diffuse ion component
observed in the upstream
\citep[e.g.][]{1981ApJ...244..711E,1982JGR....87.5063L}. While in situ
observations of energetic ions associated with interplanetary shocks and
planetary bow shocks can be well explained by the DSA theory
\citep[e.g.][]{1980JGR....85.4602S,1981JGR....86..547G}, energetic
electrons thought to be accelerated by DSA process are rarely observed
\citep{1999Ap&SS.264..481S}. On the other hand, there is no doubt about
the existence of ultra-relativistic electrons which may be accelerated
by DSA process at SNR shocks. We still have no clear consensus on what
determines the electron acceleration efficiency. The physics of electron
acceleration at collisionless shocks is poorly understood so far.

The well-known difficulty is that thermal electrons cannot be easily
scattered by \Alfven waves because of their small gyroradii. Injection
from thermal pool to mildly relativistic energy by some other mechanisms
is required. \cite{1992ApJ...401...73L} has examined electron injection
process at strictly parallel shocks. He considered the self-consistent
excitation of the whistler waves with cosmic ray electrons by applying
the standard quasi-linear theory. It was shown that the injection of low
energy electrons by the self-generated whistlers may be possible when
the \Alfven Mach number exceeds $43/\sqrt{\beta_e}$, where $\beta_{e}$
is the ratio of thermal electron pressure to magnetic pressure.  On the
other hand, \cite{1988Ap&SS.144..535P} has taken a different
approach. He considered an electron energization process within the
shock transition region by strong plasma microinstabilities. It is well
known that the reflection of upstream ions plays a dominant role in the
structure of quasi-perpendicular shocks
\citep[e.g.][]{1982JGR....87.5081L}. The reflected ions streaming
relative to the upstream plasma could excite various plasma
instabilities in the so-called foot region. \cite{1988Ap&SS.144..535P}
argued that the relative drift velocity exceeds the electron thermal
velocity at high Mach number shocks and leads to the excitation of the
Buneman instability \citep{1958PhRvL...1....8B}. The Buneman instability
gives rise to very rapid electron heating. As a result, the interaction
between incoming/reflected ions and the preheated electrons permits the
excitation of the ion acoustic instability. \cite{1988ApJ...329L..29C}
demonstrated the idea of the electron energization process by using a
hybrid simulation code where ions are treated as particles whereas
electrons are assumed to be a massless charge-neutralizing
fluid. \cite{2000ApJ...543L..67S} extended their studies to include the
electron dynamics in the shock structure by using a particle-in-cell
(PIC) code where both ions and electrons are treated as particles. They
found that localized large amplitude electrostatic solitary waves (ESWs)
are produced in the nonlinear stage of the Buneman instability. They
also argued that the rapid electron heating and acceleration are
involved with ESWs. Now, the acceleration mechanism associated with ESWs
is considered as electron shock surfing/surfatron acceleration (SSA)
process \citep{1983PhRvL..51..392K}.

SSA mechanism for ions has been extensively studied by many authors
\citep[e.g.][]{1966RvPP....4...23S,1996JGR...101.4777L,1996JGR...101..457Z}. The
process utilizes the electrostatic shock potential which is caused by
inertia difference between ions and electrons in the shock transition
region. Ions having energy smaller than the shock potential are
reflected by the shock front and begin to gyrate around the upstream
magnetic field. During their gyromotion in the upstream, they are
accelerated parallel to the shock surface by the motional electric
field. If the spatial scale of the shock potential is small compared to
the ion inertial length, multiple reflection can occur
\citep{1996JGR...101..457Z}. In contrast to this, electrons cannot be
reflected by the shock potential. The electric field directed to the
downstream is required. \cite{2001PThPS.143..149H} argued that ESWs can
play the similar role to the shock potential in ion shock surfing
acceleration, because ESWs are associated with phase space electron
holes (positively charged structures). Electrons trapped in ESWs can be
accelerated by the motional electric field. This acceleration mechanism
is very efficient, so that mildly relativistic electrons are generated
within the shock transition region on a very short time scale. The
process has attracted a considerable attention and investigated in
detail by many authors
\citep[e.g.][]{2000A&A...356..377D,2001PhRvL..87y5002M,2002ApJ...572..880H,2002ApJ...579..327S,2002ApJ...570..637S},
because it may provide a clue to the electron injection problem.

In situ observations of the Earth's bow shock also evidence the
existence of ESWs in the shock transition region and associated electron
heating
\citep[e.g.][]{1998GeoRL..25.2929B,2002ApJ...575L..25B,2006GeoRL..3315104H}.
\cite{Oka-PhD} has recently carried out a detailed investigation of the
well-resolved high Mach number Earth's bow shock ($M_A \simeq 14$)
observed by the Geotail satellite. He showed a clear coincidence between
nonthermal electrons and the appearance of the broad band electrostatic
noise (BEN) which is considered to be a signature of ESWs, although a
plausible acceleration mechanism remains unanswered.

Most of previous studies concerning electron energization processes by
microinstabilities at high Mach number shocks are restricted to the case
of strictly perpendicular shock. On the other hand, a number of PIC
simulations of quasi-perpendicular supercritical shocks have been
conducted for many decades
\citep[e.g.][]{1984JGR....89.2142F,1992PhFlB...4.3533L}. However, the
main applications of these studies were planetary bow shocks and
interplanetary shocks in the heliosphere. Here we extend these studies
to very high Mach number quasi-perpendicular shocks. We find that SSA
produces suprathermal electrons in the transition region of
quasi-perpendicular shocks. The electron energization via SSA occurs
within a relatively narrow region at the leading edge of the shock
transition region where strong electrostatic waves are observed as in
the case of strictly perpendicular shocks. The difference is that the
preaccelerated electrons are further accelerated by the so-called fast
Fermi acceleration process which is proposed by
\cite{1984JGR....89.8857W} and \cite{1984AnGeo...2..449L}. They
considered a particle motion in the de Hoffman-Teller frame (HTF) where
the motional electric field vanishes. In that frame, a particle having
sufficiently large energy is reflected by the shock which acts as a fast
moving magnetic mirror. The mirror reflection is adiabatic process,
provided that a particle gyroradius is much smaller than the shock
thickness. Therefore, the energy of reflected particle is conserved in
the HTF. Then, after the interaction with the shock, the particle
momentum parallel to the magnetic field measured in the upstream frame
is increased by $\Delta p = 2 m V_1 / \cos \theta_{Bn}$, where $m$,
$V_1$ and $\theta_{Bn}$ are respectively, particle mass, the upstream
plasma bulk velocity and the shock angle. Since the momentum increase is
proportional to the reciprocal of $\cos \theta_{Bn}$, the acceleration
becomes extremely efficient at nearly perpendicular shocks. While the
acceleration efficiency increases with increasing the shock angle, the
number of reflected particle rapidly decreases, because the initial
energy required for the reflection increases.

Later, \cite{1989JGR....9415367K} has shown that the fast Fermi process
in the HTF is equivalent to shock drift acceleration (SDA) in the normal
incidence frame (NIF) where the upstream velocity is parallel to the
shock normal. In the NIF, a particle gains its energy by drifting
parallel to the upstream motional electric field direction. Note that
\cite{1989JGR....9415367K} called the acceleration mechanism as gradient
drift acceleration and distinguished it from SDA. Since SDA was usually
based on the approximation that a particle gyroradius is large compared
to the shock width and a particle has multiple interaction with the
shock front. In this paper, however, we will not discriminate the
difference and simply call the acceleration mechanism as SDA, because
the physical mechanism is the same (gradient drift provides the energy
gain).

The process has been extensively studied in order to account for
observed energetic electrons in the upstream of the Earth's bow shocks
\citep[e.g.][]{1989JGR....9415089K,1995JGR...10021613V}. However,
detailed parametric survey of \cite{2001JGR...106.1859V} has
demonstrated quantitative discrepancies between the theoretical
expectation and observations. He argued that the process should be
modified by some other nonadiabatic processes such as pitch angle
scattering in order to explain observations. In addition to this, the
required energy for the reflection of thermal electrons at very high
Mach number shocks is unrealistically high. This process by itself
cannot account for the observed nonthermal electrons at high Mach number
shocks.

Previous studies did not consider the effect of microturbulence in the
shock transition region, which we show plays an important role in the
generation of energetic electrons. In fact, SSA is a highly efficient
acceleration mechanism, so that the preaccelerated electrons gain
sufficient energy required for the reflection process. As a result, the
preaccelerated electrons are further accelerated via SDA. This two-step
acceleration mechanism is important for providing a seed population for
subsequent DSA process. In fact, we will show that the reflected
electron energy is large enough to be accelerated by DSA process when
the Mach number is typical of SNR shocks.

We propose a theoretical model of the two-step acceleration mechanism
based on the simulation results. The present model well explains the
observed injection efficiency and the energy density of cosmic ray
electrons which were obtained by detailed analysis of \Chandra X-ray
observation of SN 1006 \citep{2003ApJ...589..827B}. Moreover, the
present model predicts the shock angle dependence of the injection
efficiency. This dependence again agrees well with the shock angle
constraint, which is required to account for the observation by the DSA
theory \citep{2003ApJ...589..827B,2004A&A...416..595Y}.

\section{SIMULATION}
We study the dynamics of ions and electrons in the self-consistent shock
structure by utilizing a one-dimensional electromagnetic PIC code where
both ions and electrons are treated as particles. A high-speed plasma
consisting of electrons and ions is injected from the left-hand boundary
of a one-dimensional simulation system and travels toward the positive
$x$. The plasma carries a uniform magnetic field $B_x$ and $B_z$. At the
right-hand boundary, the particles are specularly reflected. Then, a
shock wave is formed and propagates in the negative $x$ direction. The
downstream bulk speed becomes zero on average in the simulation
frame. Initially, there are 100 particles for each species in each
computational cell. The grid size is comparable to Debye length and the
simulation box consists of 51200 grids. The plasma parameters are as
follows: The upstream plasma $\beta_{e} = \beta_{i} = 0.08$ $(\beta_j
\equiv 8 \pi n T_j / B^2)$, where $n$, $T_j$, $B$ are the density,
temperature, and magnetic field strength, respectively. The ratio of the
plasma frequency to the electron cyclotron frequency is $\wpe/\wce =
20$. In order to reduce the computational costs, the ratio of ion to
electron mass $m_i/m_e = 100$ is used. The upstream \Alfven speed
becomes $5 \times 10^{-3} c$, where $c$ is the light speed. We use the
plasma injection four velocity of $U_0 = 5 \times 10^{-2} c$. The
\Alfven Mach number of the resultant shock wave is $M_A \simeq 15$ in
the shock rest frame. Several runs with different shock angles are
conducted with keeping the upstream magnetic field strength
unchanged. In this section, we mainly discuss the results of a run with
$\theta_{Bn} = 80 \degr$.

\begin{figure}
 \figurenum{1} \epsscale{1.0} \plotone{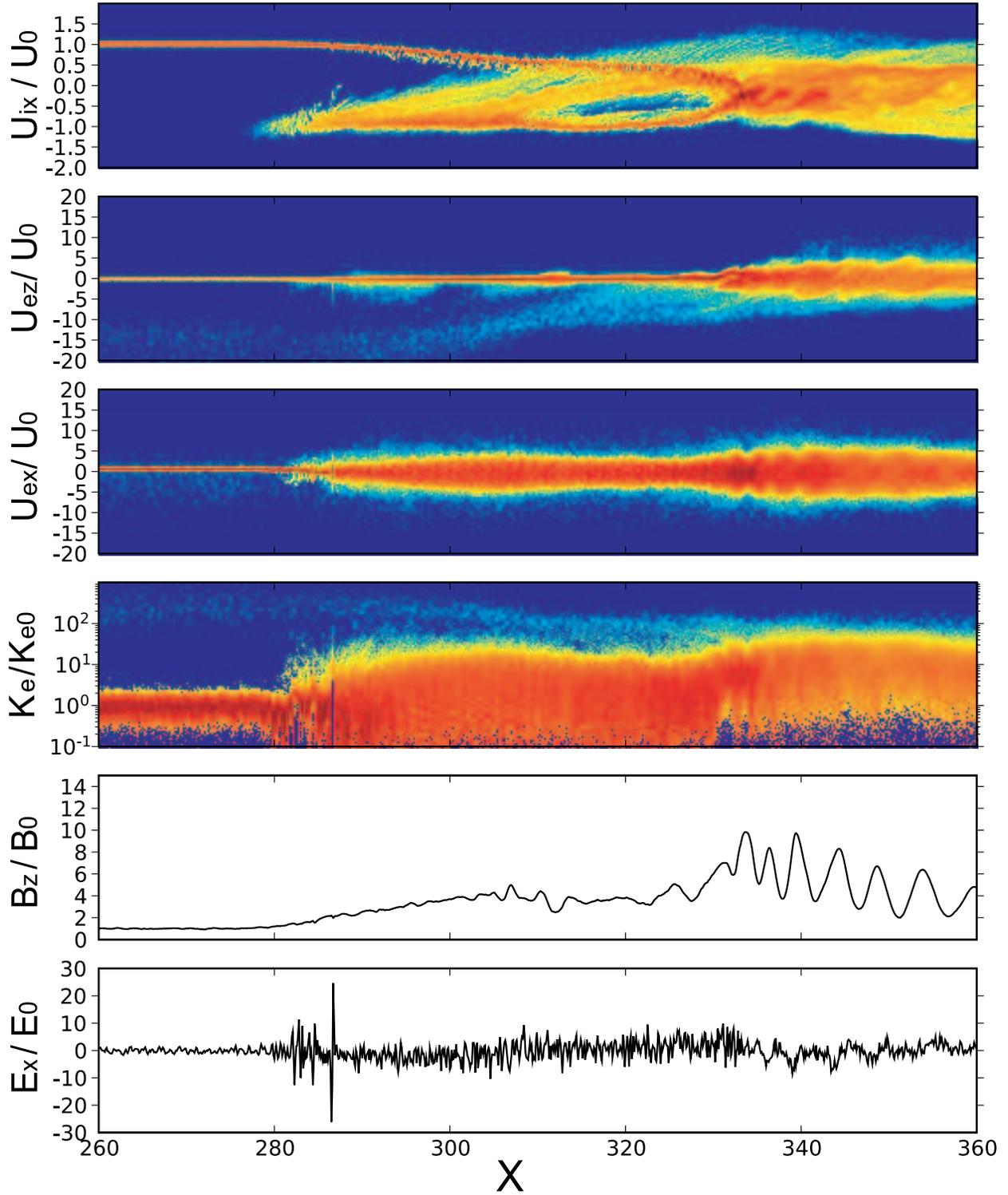} \caption{Overall shock
 structure of high Mach number quasi-perpendicular shock ($\theta_{Bn} =
 80 \degr$). Color represents the logarithm of particle count in each
 bin.} \label{fig:overall}
\end{figure}

\figref{fig:overall} shows an overall structure of the shock transition
region at $\wpe t = 12000$ (corresponding to $\wci t = 5.5$). From the
top panel, ion phase space diagram in $(X, U_{ix})$, electron phase
phase space diagram in $(X, U_{ez})$, $(X, U_{ex})$ and $(X, K_e)$, the
magnetic field $B_z$ and the electric field $E_x$, respectively. The
plasma four velocity is normalized to the injection velocity $U_0$. The
electron kinetic energy $K_e = (\gamma_e - 1) m_e c^2$ is normalized to
the injection energy $K_{e0} = (\gamma_0-1) m_e c^2$, where $\gamma_e$
and $\gamma_0$ are the Lorentz factors of each particle and the
injection velocity, respectively. The magnetic field and the electric
field are normalized to the $z$ component of upstream magnetic field
$B_{0z} = B_0 \sin \theta_{Bn}$ and the corresponding motional electric
field $E_{0y} = U_0 B_{0z} / \gamma_0 c$, respectively. The spatial
scale is given in units of the electron inertial length $c/\wpe$ in the
upstream. The color of the phase space diagrams represents the logarithm
of particle count in each bin. Note that the vertical scale of the
fourth panel (electron energy spectra) is also shown in a logarithmic
scale.

\begin{figure}
 \figurenum{2} \epsscale{1.0} \plotone{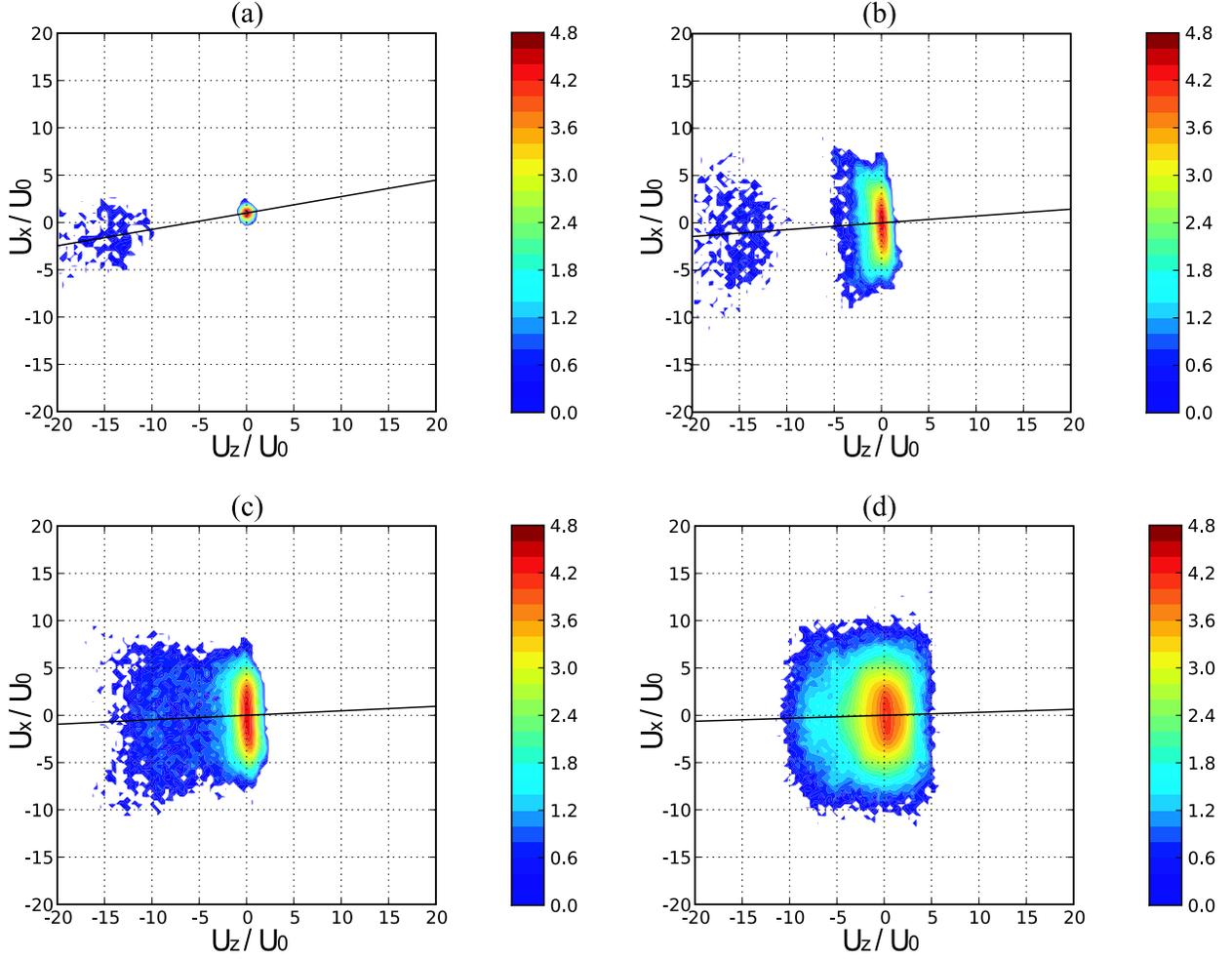} \caption{Distribution
 functions of electron in $(U_x, U_z)$ plane. Each panel shows the
 distribution function taken at (a) $260 < X < 280$, (b) $280 < X <
 300$, (c) $300 < X < 320$, (d) $320 < X < 340$, respectively. The solid
 line represents the direction of the averaged magnetic field. Color
 represents the logarithm of particle count in each bin.}
 \label{fig:psdall}
\end{figure}

\begin{figure}
 \figurenum{3} \epsscale{1.0} \plotone{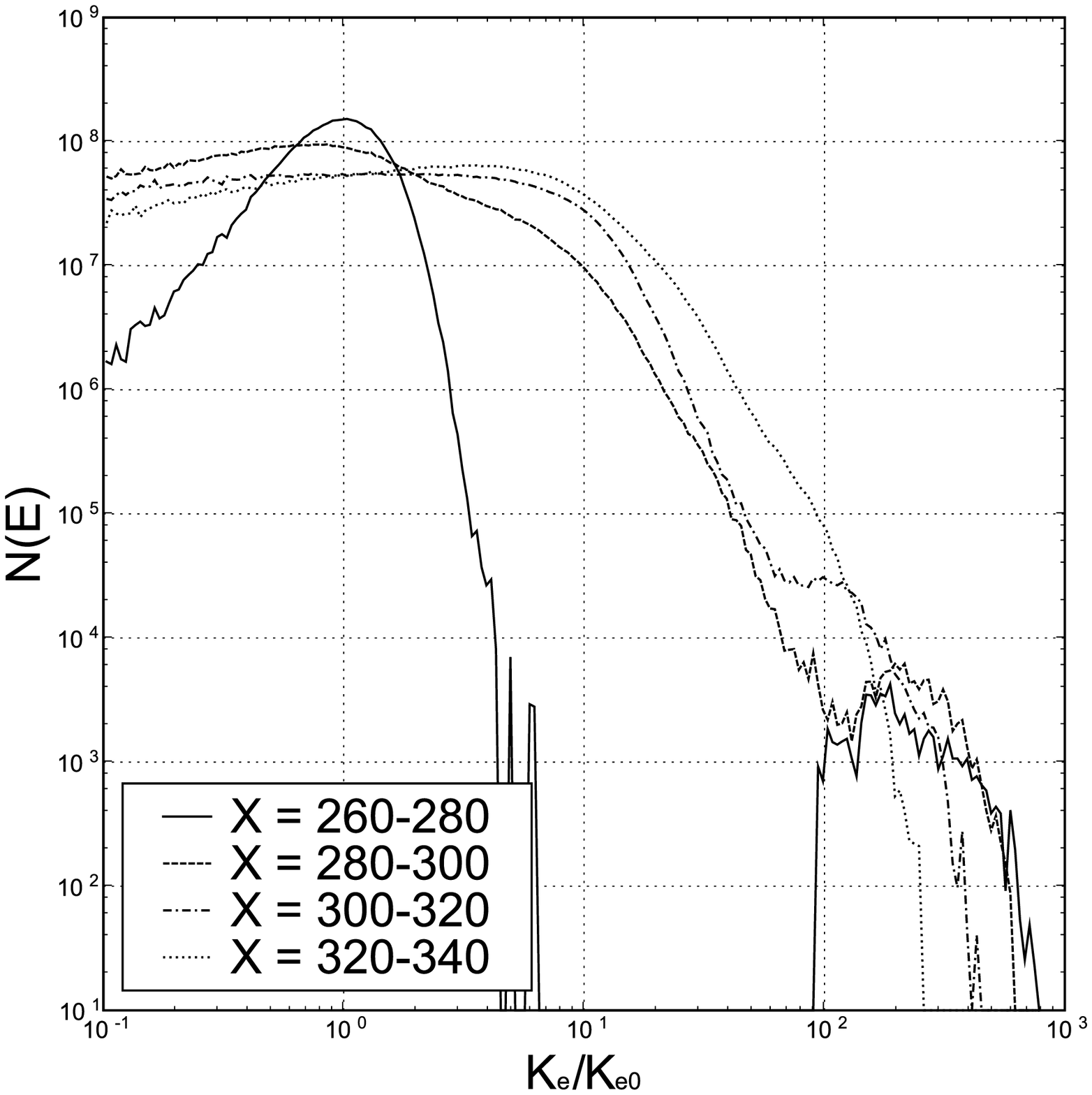} \caption{Energy spectra
 of electrons obtained around the shock transition region. Electron
 energy is normalized to the upstream bulk energy.}
 \label{fig:energy_spectra}
\end{figure}

The basic structure of the shock transition region is similar to those
obtained by previous simulation studies of strictly perpendicular shocks
\citep[e.g.][]{2000ApJ...543L..67S,2002ApJ...572..880H,2002ApJ...579..327S,2002ApJ...570..637S}. We
can find two distinct ion components, the incoming and the reflected
ions. Strong electrostatic turbulence is observed at the leading edge of
the shock transition region ($280 \lesssim X \lesssim 290$), where the
relative drift velocity between the incoming electrons and the reflected
ions becomes maximum. These electrostatic waves are excited by the
Buneman instability. Strong energization of the upstream electrons
coincides with the turbulent electrostatic waves as in the case of
strictly perpendicular shocks. The heating and acceleration of electrons
due to the turbulence occur on a very fast time scale. If we look at the
deeper inside the shock transition region, the preheated electrons
trigger the ion acoustic instability and the associated heating of
incoming ions is evident in the top panel. The maximum growth rate of
the ion acoustic instability is $\gtrsim 10 \wci^{-1}$, which we obtain
by using the standard linear dispersion analysis. The growth rate is
large enough for the instability to develop within the shock transition
region. Electrons are slowly heated up by the ion acoustic waves and the
adiabatic heating process.

In addition to these features, which are common to strictly
perpendicular shocks, we can clearly find energetic electrons streaming
away from the shock front along the magnetic field. These parallel
escaping energetic electrons can be seen in the second panel, which
represents the $z$ component of electron four velocity. Note that the
$z$ component of velocity is almost parallel to the magnetic
field. \figref{fig:psdall} displays the distribution functions of
electrons in $(U_x, U_z)$ plane taken at four different locations around
the shock transition region. Just before the shock front (a), energetic
electrons are observed as a distinct beam component. The typical beam
drift velocity parallel to the magnetic field is $u_{\parallel}/U_0 \sim
15$. Since the electron beam excites Langmuir waves in the upstream, the
cold upstream electron component is slightly modified from the far
upstream condition. With increasing penetration into the shock, the
incoming electrons are accelerated/heated up mainly perpendicular to the
magnetic field, while the parallel drift velocity of the energetic
electron beam decreases. Eventually, these two components merge into a
single, but non-Maxwellian distribution (d). \figref{fig:energy_spectra}
shows energy spectra of electrons. In the transition region ($280 < X <
340$), the middle energy range ($10 \lesssim K_e/K_{e0} \lesssim 100$)
of the spectra can be approximated by the power law with indices of
$3-4$. On the other hand, humps are found in the high energy part
($K_e/K_{e0} \gtrsim 100$) of the spectra. We observe the humps of the
energy spectra only in the upstream or at the leading edge of the
transition region ($260 < X < 300$) but not in the downstream or the
so-called overshoot region where the magnetic field strength is the
largest. The fact indicates that the observed energetic electrons are
the result of reflection (not the leakage of downstream particles). The
reflected electrons are accelerated parallel to the magnetic field
during the reflection.

\begin{figure}
 \figurenum{4} \epsscale{1.0} \plotone{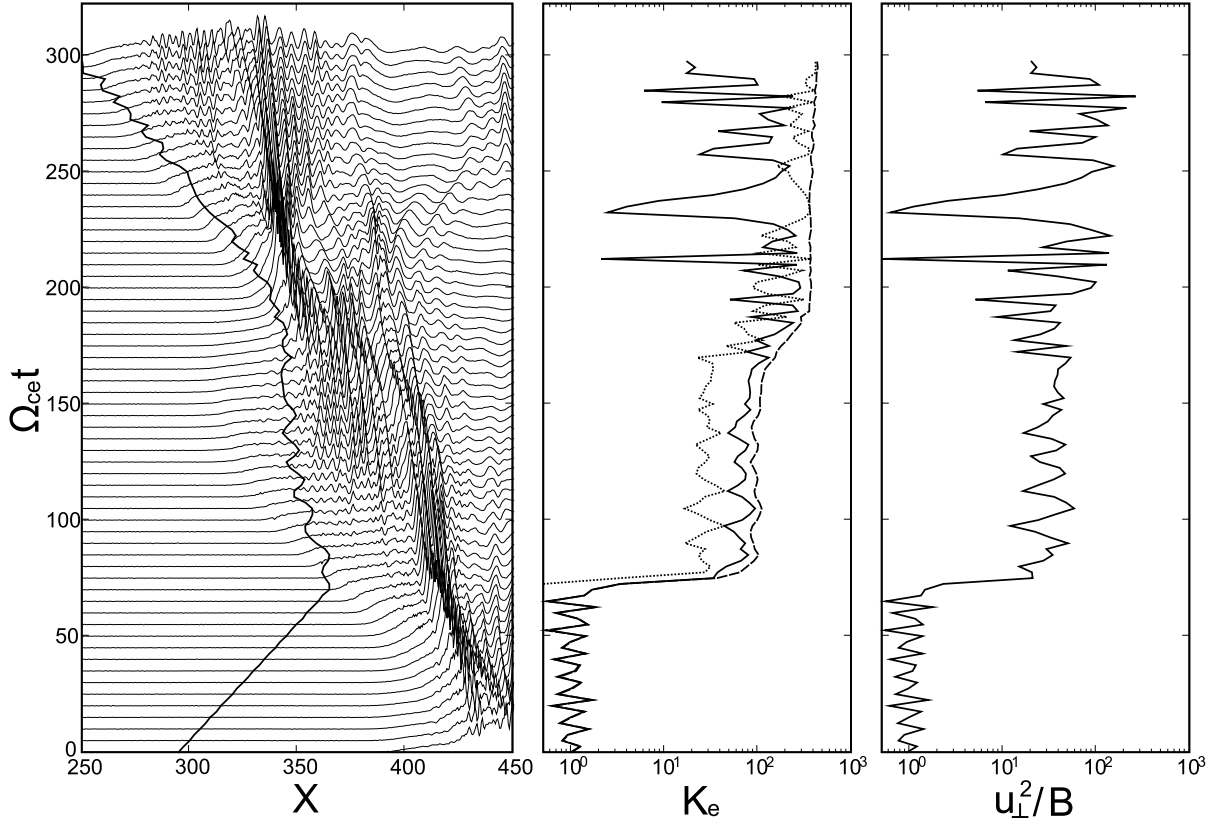} \caption{Time history of
 typical energetic particle. From left to right, particle trajectory
 (thick line) and staked profiles of the magnetic field $B_z$ (thin
 lines), particle energy, the first adiabatic invariant,
 respectively. The solid, dotted and dashed lines of the middle panel
 show the perpendicular, parallel and total particle energy,
 respectively. Energy and the first adiabatic invariant are normalized
 to the upstream values.}  \label{fig:trajectory}
\end{figure}

In order to understand the electron acceleration process in more detail,
individual trajectories of energetic electrons are
analyzed. \figref{fig:trajectory} shows a trajectory of typical
energetic electron. The left panel represents the trajectory of electron
(thick line) with the stacked profiles of the magnetic field $B_z$ (thin
lines). The middle panel shows the time history of the particle
energy. The solid, dotted and dashed lines of the panel display the
perpendicular, parallel and total energy of the particle,
respectively. The right-hand panel shows the time history of the first
adiabatic invariant normalized to the upstream value. The unit of the
vertical axis is the reciprocal of the electron gyrofrequency
$\wce^{-1}$. The time history is plotted after $\wce t = 300$ ($\wpe t =
6000$). As is evident from the stack plot of the magnetic field
profiles, the shock front is highly nonstationary and periodically
reforms itself with the characteristic time scale of $\sim 2
\wci^{-1}$. The shock wave is propagating toward the negative $x$
direction with the average speed of $\sim 0.5 U_0$. Initially, the
particle located in the upstream is convected toward the shock with the
$\V{E} \times \V{B}$ drift velocity. The particle first encounters the
shock front at $\wce t \sim 70$, and gains quickly its energy within a
short time of $\wce \Delta t \sim 5$. The energy of the particle
increases by 2 orders of magnitude during the interval and the energy
gain is almost perpendicular to the magnetic field. It is evident from
the increase of the first adiabatic invariant that the process is highly
nonadiabatic. This is indeed SSA mechanism investigated in detail by the
previous studies \citep[e.g.][]{2002ApJ...572..880H}. The properties of
the acceleration are consistent with the previous results and we will
not discuss the details of the process in any detail. After the first
energization, the particle slowly drifts around the transition region
without changing its energy until it collides with the overshoot at
$\wce t \sim 180$. The particle is pushed toward the upstream direction
by the magnetic mirror force during the collision and gains its total
energy, whereas the first adiabatic invariant remains almost
constant. In other words, the particle is reflected by the shock acting
as a fast moving magnetic mirror and gains its momentum parallel to the
magnetic field. The parallel energy increases, while the perpendicular
energy is almost constant or slightly decreases during the interaction.

The latter acceleration process associated with the reflection is known
as SDA.  As the shock angle and the upstream bulk flow speed increase,
the energy gain of SDA increases. On the other hand, since the initial
required energy for the reflection increases, the number of reflected
particle rapidly decreases. Therefore, we expect that SDA process would
not operate at high Mach number shocks. It is readily shown that the
upstream thermal electrons of the present simulation cannot be reflected
by the shock: The bulk velocities of the plasma in the upstream and
downstream measured in the shock rest frame are $U_1/U_0 \simeq 1.5$ and
$U_2/U_0 \simeq 0.5$, respectively. The effective velocity of ``the
magnetic mirror'' toward the upstream is $U_s/U_0 = U_1/U_0 / \cos
\theta_{Bn} \simeq 8.6$. The loss cone angle denoted by $\theta_{c}$ is
given by
\begin{eqnarray}
 \sin \theta_{c} = \sqrt{ \frac{B_1}{B_{max}} }, \label{eq:loss-cone}
\end{eqnarray}
where $B_1$ and $B_{max}$ are the magnetic field strength in the
upstream and the overshoot, respectively. The reflection will take place
when the condition $u_{\perp} \gtrsim U_s \sin \theta_{c}$ is satisfied,
where $u_{\perp}$ denotes the perpendicular velocity of particle
measured in the upstream frame. If we take a typical magnetic
compression ratio of $B_{max}/B_1 \simeq 10$, the required velocity
becomes $U_s \sin \theta_{c} \simeq 2.7 U_0$. This condition is quite
severe, because the upstream electron thermal velocity is only $0.2
U_0$. Here, for simplicity, we ignore the effect of the electrostatic
shock potential which suppresses the reflection efficiency. However, it
cannot be neglected in general; hence, the reflection of thermal
electrons requires even more stringent condition.

\begin{figure}
 \figurenum{5} \epsscale{1.0} \plotone{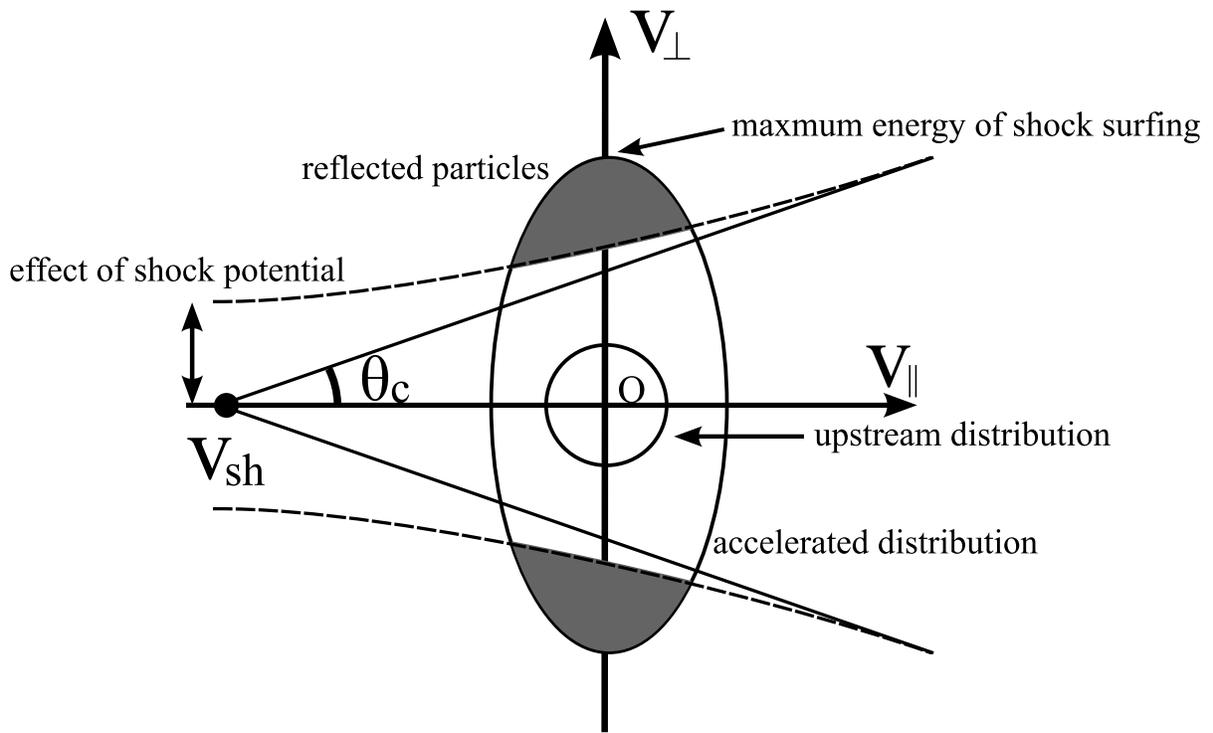} \caption{Schematic
 illustration of surfing and drift acceleration.}
 \label{fig:surfing-drift}
\end{figure}

In contrast to the above theoretical analysis without any nonadiabatic
process, our simulation results demonstrate the reflection does indeed
take place, which we attribute to the presence of the preacceleration
via SSA. \figref{fig:surfing-drift} shows a schematic illustration of
the reflection process initiated by SSA process. The cold upstream
electrons are energized by the electrostatic waves excited by the
Buneman instability at the leading edge of the shock transition
region. The energization is so efficient that a nonnegligible fraction
of electrons escapes outside the loss cone on a time scale of $\sim
\wce^{-1}$. The preaccelerated electrons escaping from the loss cone are
subjected to SDA. In fact, by back tracing the trajectories of energetic
electrons observed in the upstream side, we can confirm the scenario,
i.e., the reflected electrons suffer the rapid energization via SSA when
they enter the shock transition region and are reflected by the
overshoot.  On the other hand, low energy electrons are never reflected
and just transmitted to the downstream. Therefore, we regard the
two-step acceleration process as a preferential reflection process of
energetic electrons. In the present mechanism, SSA plays a crucial role
to trigger the secondary acceleration process. The important point is
that SSA produces suprathermal particles with approximately the power
law energy spectra. Since the shock potential considerably reduces the
reflection efficiency, the expected heating due to the Buneman
instability is not sufficient to provide the required reflection energy
and the production of suprathermal particles is essential. It is also
important that the acceleration is almost perpendicular to the magnetic
field. Because of this, pitch angles of energetic electrons become large
and they are easily reflected by the shock.

\section{ELECTRON INJECTION MODEL}
The reflection of energetic electrons discussed in the previous section
can be considered as ``electron injection'' to subsequent DSA
process. The reflected electrons observed as a beam component in the
upstream drive several instabilities. It is well known that fast
electron stream parallel to the magnetic field excites Langmuir waves
via the bump-on-tail instability. We observe the enhanced Langmuir
turbulence in the upstream of the simulation results, which is not seen
in strictly perpendicular shocks. Energetic electrons streaming away
from the shock and associated electrostatic turbulence are observed in
the foreshock region of the Earth's bow shock
\citep[e.g.][]{1969JGR....74...95A,1981JGR....86.4493A,2000JGR...105...79K}. Another
electromagnetic instability may also be excited by electron cyclotron
resonance. We expect a left-hand polarized electromagnetic wave
propagating parallel to the beam which can scatter the energetic
electron themselves, provided that the beam speed is greater than $\sim
v_{A} \frac{m_i}{m_e}$. As we will see in the discussion, this condition
can be satisfied at shocks with Mach number typical of SNR
shocks. Because of this, the initial field-aligned beam will tend to
relax and become isotropic. From the above consideration, we recognize
the reflected electrons as the seed population of DSA process, although
a nonlinear evolution of the energetic electrons cannot be followed by
our simulations because of the use of one-dimensional simulation box and
a limited system size.

The density of reflected electrons is identical to the injection
efficiency defined as $n_e^{NT}/n_e$, where $n_e^{NT}$ is the number of
nonthermal particles. Therefore, it is important to construct a
theoretical model of the two-step acceleration process which predicts
the injection efficiency and the energy density of nonthermal electrons
relative to the thermal energy density.

\subsection{Electron Heating and Acceleration in the Foot}
We investigate very high Mach number quasi-perpendicular shocks where
the excitation of the Buneman instability is expected at the leading
edge of the shock transition region. These shocks are typically observed
around young SNRs. If we use the typical values of the magnetic field
$\sim 10 \mug$, the density $\sim 0.1 \rm{cm}^{-3}$ and the shock speed
$\sim 10^4 \kms$ \cite[e.g.][]{2003ApJ...589..827B}, the Mach number of
the shock becomes $M_A = 100-1000$. Although temperatures of upstream
plasmas are difficult to estimate, they are probably very cold ($\beta_e
< 1$). Therefore, we can expect that the threshold condition of the
Buneman instability
\begin{eqnarray}
 M_A \gtrsim \sqrt{\beta_e \frac{m_i}{m_e}} \label{eq:buneman}
\end{eqnarray}
is satisfied at these shocks.

In the following discussion, we neglect the relativistic effect for
simplicity. However, we will show that the model agrees well with the
simulation results which are in weakly relativistic regime (the maximum
Lorentz factor of energetic electrons is $\gamma_e \sim 2$).

Since the free energy of the Buneman instability is the relative
streaming between the incoming electrons and the reflected ions, the
amplitude of the electric field can be estimated as
\begin{eqnarray}
 \alpha \frac{1}{2} n_e m_e V_d^2 \simeq \frac{E^2}{8 \pi},
\end{eqnarray}
where $\alpha$ is a energy conversion factor and $V_d$ is the relative
drift velocity (typically $V_d \sim 2 V_1$). Although the precise value
of $\alpha$ is hard to determine, it is a factor of the order of unity
\citep{1981PhFl...24..452I,2000PhPl....7.5171D}. We assume that the wave
electric field is parallel to the shock normal as in the case of our
one-dimensional simulations. Thus, the electric fields parallel and
perpendicular to the upstream magnetic field are respectively given by
$E_{\parallel} = E \cos \theta_{Bn}$ and $E_{\perp} = E \sin
\theta_{Bn}$, where $\theta_{Bn}$ is the shock angle. By equating the
electric field energy to the electron thermal energy in each direction
(parallel and perpendicular), we estimate the resultant electron thermal
velocity $v_{e,\perp}$, $v_{e,\parallel}$ after saturation as
\begin{eqnarray}
 v_{e, \perp} &\simeq& V_1 \sin \theta_{Bn} \\ v_{e, \parallel} &\simeq&
 V_1 \cos \theta_{Bn},
\end{eqnarray}
respectively \citep{1988Ap&SS.144..535P}.

In addition to the strong electron heating, the simulation results
clearly demonstrate the formation of high energy tail in the electron
energy spectra due to the Buneman instability. Therefore, we employ the
bi-kappa distribution (\ref{eq:bi-kappa}) as a model distribution
function in the foot,
\begin{eqnarray}
 f( v_{\parallel}, v_{\perp} ) = \frac{n_{\foot}}{v_{e,\perp}^2
  v_{e,\parallel}} \frac{ \Gamma(\kappa+1) }{\left(\pi
  \kappa\right)^{3/2} \Gamma(\kappa-1/2)} \left[ 1 + \frac{1}{\kappa}
  \left( \frac{v_{\perp}^2}{v_{e,\perp}^2} + \frac{(v_{\parallel} -
  \Vsh)^2}{v_{e,\parallel}^2} \right) \right]^{-\kappa-1}
  \label{eq:bi-kappa}
\end{eqnarray}
where $\Gamma(x)$ is the gamma function, $\Vsh$ and $n_{\foot}$ are the
parallel drift velocity and the density in the foot, respectively. The
distribution function is measured in the HTF. We define the foot as the
region where the Buneman instability saturates. Hereafter, the subscript
$f$ represents the value in the foot. The incoming electrons are
decelerated so as to cancel the zeroth order current, and somewhat
compressed ($n_{\foot} > n_1$). The mass conservation law leads to
$V_{\foot} = V_1 n_1 / n_{\foot}$. The magnetic field is also compressed
with the same compression ratio $B_{\foot} = B_{1} n_{\foot} /
n_{1}$. Thus, the parallel drift velocity of the compressed incoming
electron is given by
\begin{eqnarray}
 \Vsh = \frac{ V_{\foot} }{ \cos \theta_{B_{\foot} n} },
\end{eqnarray}
which is almost equal to $V_1 / \cos \theta_{Bn}$ for $\theta_{Bn}
\simeq 90 \degr$. Here $\theta_{B_{\foot} n}$ is defined as the angle
between the shock normal and the magnetic field line in the foot.

\subsection{Adiabatic Reflection}
After the energization due to the Buneman instability, the ion acoustic
instability is triggered and further electron heating may
occur. However, this heating is not important for the generation of
suprathermal electrons. Since we find that the first adiabatic invariant
of energetic electron is conserved during the reflection process, we use
the adiabatic approximation after the energization due to the Buneman
instability. This approximation enables us to compute the reflected
electron density by integrating the distribution function outside the
loss cone. In general, the electrostatic shock potential is known to
affect the mirror reflection process. If we include the finite potential
measured in the HTF $\phi^{HT}$, the condition of the electron
reflection becomes
\begin{eqnarray}
 v_{\perp}^2 \ge \left( v_{\parallel}^2 + \frac{2 e}{m_e}\phi^{HT}
  \right) \tan \theta_c
\end{eqnarray}
where $\theta_c$ is the loss cone angle given by
\begin{eqnarray}
 \sin \theta_c = \sqrt{ \frac{B_{\foot}}{B_{\overshoot}} }.
\end{eqnarray}
Here we use the foot to the maximum compression ratio to estimate the
loss cone angle, because we use the adiabatic approximation after the
energization in the foot. We consider $B_{\overshoot}$ as the magnetic
field strength in the overshoot. We know from both observations and
numerical simulations that the shock potential is determined by the
upstream bulk ion flow energy. Therefore, we normalize the shock
potential to the bulk ion energy as
\begin{eqnarray}
 \phiht = \frac{ e {\phi^{HT}} }{ \frac{1}{2} m_i V_1^2 }.
   \label{eq:potential}
\end{eqnarray}
The effect of the potential cannot be neglected when the particle
parallel energy is comparable or smaller than the potential. Since
the typical parallel velocity is expressed as $v_{\parallel} \simeq \Vsh
\simeq V_1/\cos \theta_{Bn}$, the condition can be written as
\begin{eqnarray}
 \phiht \frac{m_i}{m_e} \cos^2 \theta_{Bn} \gtrsim 1.
\end{eqnarray}
For $\phiht = 0.4$ and $m_i/m_e = 100$, this condition leads to
$\theta_{Bn} \lesssim 81 \degr$. Moreover, this critical angle increases
with increasing $m_i/m_e$. The use of the realistic proton to electron
mass ratio gives $\theta_{Bn} \lesssim 88 \degr$. Consequently, the
effect of the shock potential is important for a wide range of shock
angles and should be included in the injection model. We should note
that the use of unrealistically small value of $m_i/m_e$, which is usual
in most of PIC simulations, overestimates the reflection
efficiency.

Although the shock potential in the HTF is identical to that in the NIF
in the absence of the noncoplanar magnetic field component, significant
field rotations out of the coplanarity plane within the shock transition
region are often found by both in situ observations and numerical
simulations \citep[e.g.][]{1987JGR....92.2305T}. Our numerical
simulations also produce the noncoplanar magnetic field component
($B_y$) in the transition region. However, \cite{1989JGR....9415367K}
showed that the effect of the noncoplanar magnetic field component is
only of second order in $B_y/B$ and thus has little influence on the
electron kinetics. Therefore, we can use the potential measured in the
NIF as $\phi^{HT}$.

The reflected electron density $n_r$ can be written as
\begin{eqnarray}
 n_r = 2 \pi \int_{0}^{\infty} d v_{\parallel} \int_{
  \sqrt{v_{\parallel}^2 + \frac{2 e}{m_e} \phi^{HT} } \tan \theta_c
  }^{\infty} v_{\perp} d v_{\perp} f(v_{\parallel}, v_{\perp}).
\end{eqnarray}
Evaluating the integral analytically, we obtain
\begin{eqnarray}
 n_r = \frac{n_{\foot}}{2 v_{e,\parallel} r^{\kappa} \sqrt{\kappa \pi}}
  \left[ \sqrt{\frac{r}{p} \pi} - \frac{2 q}{p}
  \frac{\Gamma(\kappa)}{\Gamma(\kappa-1/2)} F(\frac{1}{2}, \kappa,
  \frac{3}{2}; -\frac{q^2}{r p}) \right], \label{eq:beam-density}
\end{eqnarray}
where $F(\alpha, \beta, \gamma; x)$ is the Gauss hypergeometric
function and $p$, $q$, and $r$ are respectively given by
\begin{eqnarray}
 p &=& \frac{\tau + \tan^2 \theta_c}{\kappa v_{e,\perp}^2} \\
 q &=& \frac{\Vsh}{\kappa v_{e,\parallel}^2} \\
 r &=& \frac{\frac{2 e}{m_e} \phi^{HT} \tan^2 \theta_c}{\kappa v_{e,\perp}^2}
  + \frac{\Vsh^2}{\kappa v_{e,\perp}^2}
  \frac{\tau \tan^2 \theta_c}{\tau + \tan^2 \theta_c} + 1,
\end{eqnarray}
using the anisotropy defined as $\tau \equiv v_{e,\perp}^2 /
v_{e,\parallel}^2$.

The reflected electron beam velocity measured in the shock frame can be
easily estimated from the adiabatic theory as
\begin{eqnarray}
 v_{r} = \Vsh \sqrt{ \frac{ B_{\overshoot} - B_{\foot}}{B_{\foot} -
  B_1}}.  \label{eq:beam-velocity}
\end{eqnarray}
The beam temperature can be determined by the integral
\begin{eqnarray}
 n_r T_r = 2 \pi \int_{0}^{\infty} d v_{\parallel} \int_{
  \sqrt{v_{\parallel}^2 + \frac{2 e}{m_e} \phi^{HT} } \tan \theta_c
  }^{\infty} v_{\perp} d v_{\perp} \frac{1}{2} m_e (v_{\parallel}^2 +
  v_{\perp}^2) f(v_{\parallel}, v_{\perp}). \label{eq:analytic-energy}
\end{eqnarray}
However, we use a rather simple estimate instead of this. The thermal
energy density of the reflected part of the distribution function can be
written as
\begin{eqnarray}
 E_{th} = \int_{\Emin}^{\infty} N(E) E dE = \int_{\Emin}^{\infty} A
  E^{-k} dE,
\end{eqnarray}
where $A$ and $\Emin$ are the normalization constant and the minimum
energy required for the reflection, respectively. In the above integral,
we approximate the energy spectrum by the power law part of the
distribution function and the anisotropy is neglected. Substituting the
normalization constant $A = n_r \kappa E_{\rmin}^{\kappa}$, we obtain
\begin{eqnarray}
 E_{th} = n_r \frac{\kappa}{\kappa-1} E_{\rmin}.
\end{eqnarray}
The minimum energy is approximated by
\begin{eqnarray}
 E_{\rmin} \simeq \left( \frac{1}{2} m_e \Vsh^2 + e \phi^{HT} \right)
  \tan^2 \theta_c.  \label{eq:emin}
\end{eqnarray}
By combining these equations, we can write the energy density of the
beam $\epsilon_r$ as
\begin{eqnarray}
 \epsilon_r \simeq \frac{1}{2} n_r m_e v_r^2 + \frac{\kappa}{\kappa-1}
  n_r \left( \frac{1}{2} m_e \Vsh^2 + e \phi^{HT} \right) \tan^2
  \theta_c.  \label{eq:beam-energy}
\end{eqnarray}
This is somewhat crude estimate. Since the reflected beam distribution
will be modified by nonadiabatic processes in the shock transition
region, even the energy density estimated from the analytic integration
(\ref{eq:analytic-energy}) might be inaccurate. Nonetheless,
(\ref{eq:beam-energy}) gives reasonably good estimate if it is compared
to the simulation results. Therefore, we adopt this simple
approximation. In the following discussion, density, velocity and energy
will be given in units of the density, bulk velocity and bulk electron
energy in the upstream, respectively. Note that, under this
normalization, the result of the present model is independent of $M_A$
and $\beta_e$, provided that (\ref{eq:buneman}) is satisfied.

\subsection{Comparison Between Model and Simulation Results}
Obviously, the injection efficiency strongly depends on the choice of
$\kappa$. We adopt $\kappa = 2.5$ (corresponding to the power law index
of $3.5$) which approximates the energy spectra in the shock transition
region (see \figref{fig:energy_spectra}). Likewise, the value of $\phiht
= 0.4$ is used for the model calculation. The results depend weakly on
the magnetic field in the foot and the overshoot. In the following
discussion, we use the fixed values of $B_{\foot}/B_1 = 2.5$ and
$B_{\overshoot}/B_1 = 10.0$ which are the typical values obtained from
the simulation results. The variation of these values does not affect
the result significantly.

\begin{figure}
 \figurenum{6} \epsscale{1.0} \plotone{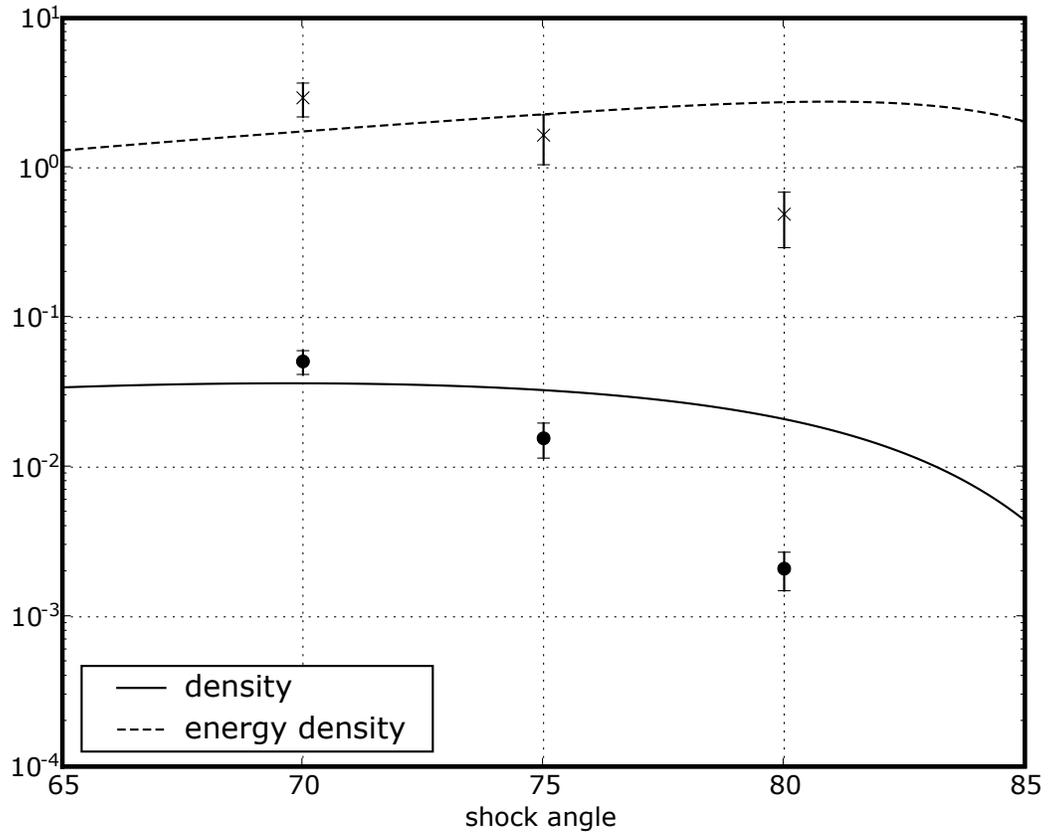} \caption{Comparison
 between our theoretical model and the simulation results. The solid and
 the dashed lines show density and energy density of reflected electrons
 calculated by the model. The simulation results are displayed by filled
 circles (density) and crosses (energy density).}
 \label{fig:uncorrected}
\end{figure}

\figref{fig:uncorrected} shows the shock angle dependence of the density
and the energy density of the reflected electrons calculated by the
model with those obtained from the simulation results. The energy
density of the model is converted to the downstream rest frame.
The simulation results are averaged over $X_{sh} - 70 c/\wpe \le x \le
X_{sh} - 20 c/\wpe$, where $X_{sh}$ is the position at which the
magnetic field is compressed by a factor of $1.5$ from the upstream
value. The time-averaged values during the last interval $\Delta T
\simeq \wci^{-1}$ of each run are shown. The error bars correspond to
the variance during the interval.

The model curves agree well with the simulation results when the shock
angle is small ($\theta_{Bn} \lesssim 75$). However, the simulation
results tend to deviate from the model curves with increasing the shock
angle.  At the shock angle of $\theta_{Bn} = 85\degr$, no reflected
electrons are observed in the simulation, while the model gives the
density of $\sim 4 \times 10^{-3}$. This large difference is not
acceptable when we apply the model to real shocks. In the next section,
we introduce some modifications into the model in order to correct the
discrepancy.

\subsection{Corrective Effects}
\subsubsection{Maximum Energy Cut-Off of Shock Surfing Acceleration}
The actual energy spectra obtained from the simulations have a cut-off
energy where the number of particle rapidly falls off. Therefore, the
number of reflected electron will fall off as $\propto \exp \left( -
\Emin / \Emax \right)$, where $\Emax$ is the maximum cut-off energy of
SSA. It is easy to understand that the presence of the maximum cut-off
energy becomes important when it is comparable or smaller than
$\Emin$. Because $\Emin$ increases with the shock angle, the presence of
the maximum energy introduces the cut-off shock angle above which the
reflection efficiency rapidly falls off. We can determine $\Emax \sim
100$ from the simulation results. This leads to the cut-off shock angle
of $\sim 86\degr$, which is larger than that of the simulation results.

Although only the above correction effect cannot explain why the
reflected electron density of the simulation results show the rapid
decrease at the shock angle $\theta_{Bn} \gtrsim 80$, the effect may be
still important even when the shock angle is smaller than the cut-off,
if we consider the application to real shocks. This is due to the fact
that $\Emin$ strongly depends on the shock potential which increases
with increasing the mass ratio. As discussed later, $\Emax$ should be
larger than the shock potential in order to obtain a large fraction of
reflected electrons.

\subsubsection{Escape Probability During Reflection}
Although we use the adiabatic approximation after the first energization
to compute the reflected electron density, the first adiabatic invariant
may be violated in the presence of nonstationarity of the shock
structure and/or the ion acoustic turbulence in the transition
region. Particles which are initially outside the loss cone may be
scattered by these effects and fall inside the loss cone. If we consider
the escape probability $\Pesc$ as a constant both in time and space
during the reflection, the number density will fall off as $\propto \exp
\left(-\Pesc \Tref \right)$, where $\Tref$ is the characteristic time
for the reflection process. We can write $\Tref$ by using initial
perpendicular particle energy $\Eref$ (see Appendix)
\begin{eqnarray}
 \wci \Tref \simeq \frac{2 m_e \Vsh^2} {\Eref/\tan^2 \theta_c - e
  \phi^{HT}},
\end{eqnarray}
where the spatial gradient of the magnetic field strength and the
electrostatic potential are assumed to be constant and the shock width
is evaluated as $V_1 / \wci$. We approximate the typical reflected
electron energy before the reflection process as $\Eref \simeq (\Emax +
\Emin)/ 2$. It is readily shown that $\Tref$ increases with increasing
the shock angle, which leads to the reduction of the reflection
efficiency. We should note that the parameter $\Pesc$ is introduced as a
free parameter to correct the discrepancy between the model and the
simulation results. We determine the value of $\Pesc$ so that the model
fits the simulation results.

Thereafter, $\Pesc$ will be given in the units of ion gyrofrequency $\wci$,
because $\Tref$ is of the order of ion gyroperiod.

\subsubsection{Comparison Between Corrected Model and Simulation Results}
Rewriting the density given by (\ref{eq:beam-density}) as
$n_r^{\prime}$, above two effects can be included in the model with the
following form
\begin{eqnarray}
 n_r = n_r^{\prime} \exp \left( - \frac{\Emin}{\Emax} - \Pesc \Tref
  \right).  \label{eq:density}
\end{eqnarray}

\begin{figure}
 \figurenum{7} \epsscale{1.0} \plotone{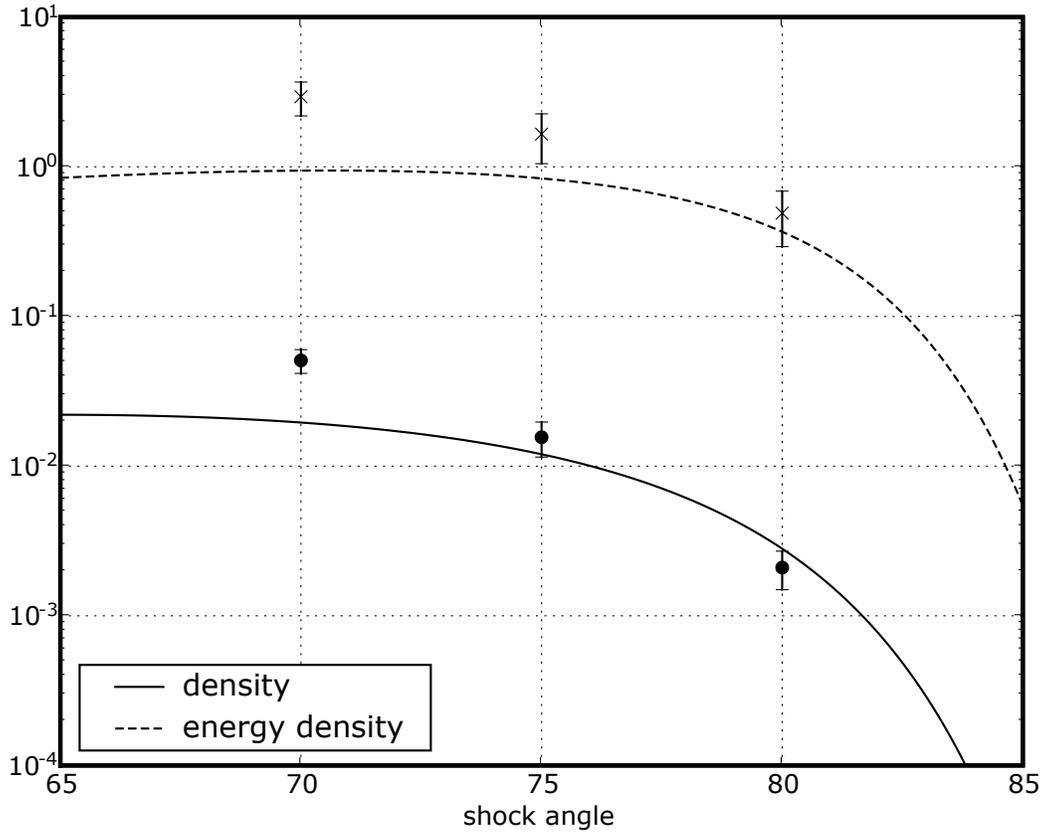} \caption{Comparison
 between our theoretical model with corrections and the simulation
 results. The format is the same as \figref{fig:uncorrected}. $\Pesc =
 4$ and $\Emax = 100$ are used to produce the figure.}
 \label{fig:corrected}
\end{figure}

\figref{fig:corrected} shows a comparison between the simulation results
and the model using the above correction. The format of the figure is
the same as that of \figref{fig:uncorrected}.  We use $\Pesc = 4$ and
$\Emax = 100$ to produce the figure. In this case, we see that the model
is significantly improved and the agreement becomes quite
good. Therefore, we can conclude that the present model can predict the
injection efficiency as well as the energy density of nonthermal
particles, if the model parameters are properly estimated.

Note that the value $\Pesc = 4$ used here indicates that the underlying
mechanism of electron scattering in the shock transition region is the
self-reformation of the shock front, which occurs with the
characteristic time scale of $\sim 2 \wci^{-1}$. Since the shock
structure considerably changes with this time scale, the motion of
electrons interacting with the shock longer than this period may also be
modified. As a result, only electrons which have sufficient energy for
the rapid reflection can escape to the upstream.

\subsection{Application to Supernova Shocks}
Now, let us apply the model to SNR shocks. First, we should discuss
the maximum energy of SSA. Since the shock potential is determined by
the ion bulk flow energy, it will increase with increasing the mass
ratio. The use of the same maximum energy $\Emax$ for the real mass
ratio shock will result in the net reduction of the injection
efficiency.

\cite{2002ApJ...572..880H} discussed the maximum energy of SSA and
obtained the condition
\begin{eqnarray}
 2 M_{A} \sqrt{\alpha \frac{m_e}{m_i} } \geq 1
\end{eqnarray}
for ``unlimited electron acceleration'', which means trapped electrons
cannot escape from ESWs and continue to accelerate. They also mentioned
that even in this regime, the trapping time might be limited by some
other important factors. These include shock front nonstationarity
(e.g. self-reformation) and multidimensional effects. In addition to
this, the stability of ESWs is also a quite important factor limiting
the trapping time. \cite{2000PhPl....7.5171D,2004PhRvL..92f5006D}
investigated the lifetime of the BGK mode excited by the Buneman
instability using a periodic simulation model. They showed that the
lifetime depends on the drift velocity of the reflected ions and also
the background ion temperature. In our current understanding, it is
still a hard task to estimate the trapping time scale. Furthermore,
\cite{2001PhRvL..87y5002M} showed that stochastic electron acceleration
can occur when $\wpe/\wce \gg 1$. In other words, electrons once
detrapped from the potential can interact with the wave again. In this
case, the maximum energy of SSA is not simply limited by the trapping
time scale. It is very difficult to estimate the maximum energy of SSA
and far beyond the scope of the present paper. We simply assume that the
maximum energy increases with increasing the mass ratio from the
following qualitative consideration: If the mass ratio is increased, the
free energy provided by the reflected ions also increases. Since the
free energy is the source of the energetic electrons, the maximum energy
of SSA will also become higher.

If the maximum energy is smaller than the shock potential, the injection
efficiency is greatly reduced independent of the shock angle. For
typical values of the shock potential $\phiht \sim 0.4$, we require
$\Emax \gtrsim 750$ in order to obtain a measurable fraction of the
reflected electrons. We assume $\Emax = 1000$ in the following
discussion, which leads to the cut-off shock angle of $\sim 88
\degr$. In this case, the presence of the maximum energy cut-off does
not significantly affect the result when the shock angle is smaller than
the cut-off shock angle. Note that $\Emax = 1000$ corresponds to $10-100
\; \kev$ at typical SNR shocks ($V_1/c \sim 10^{-2}$).

The escape probability $\Pesc$ is also important in the sense that it
introduces another cut-off shock angle. If $\Pesc$ is determined by the
frequency of the self-reformation, it should be independent of the mass
ratio. Therefore, we adopt $\Pesc = 4$ obtained by the simulation
results of $m_i/m_e = 100$. This leads to the cut-off shock angle of
$\sim 85 \degr$, which is smaller than that introduced by
$\Emax$. Namely, in the present parameter range, $\Pesc$ is more
important than $\Emax$ for determining the shock angle dependence.

\begin{figure}
 \figurenum{8} \epsscale{1.0} \plotone{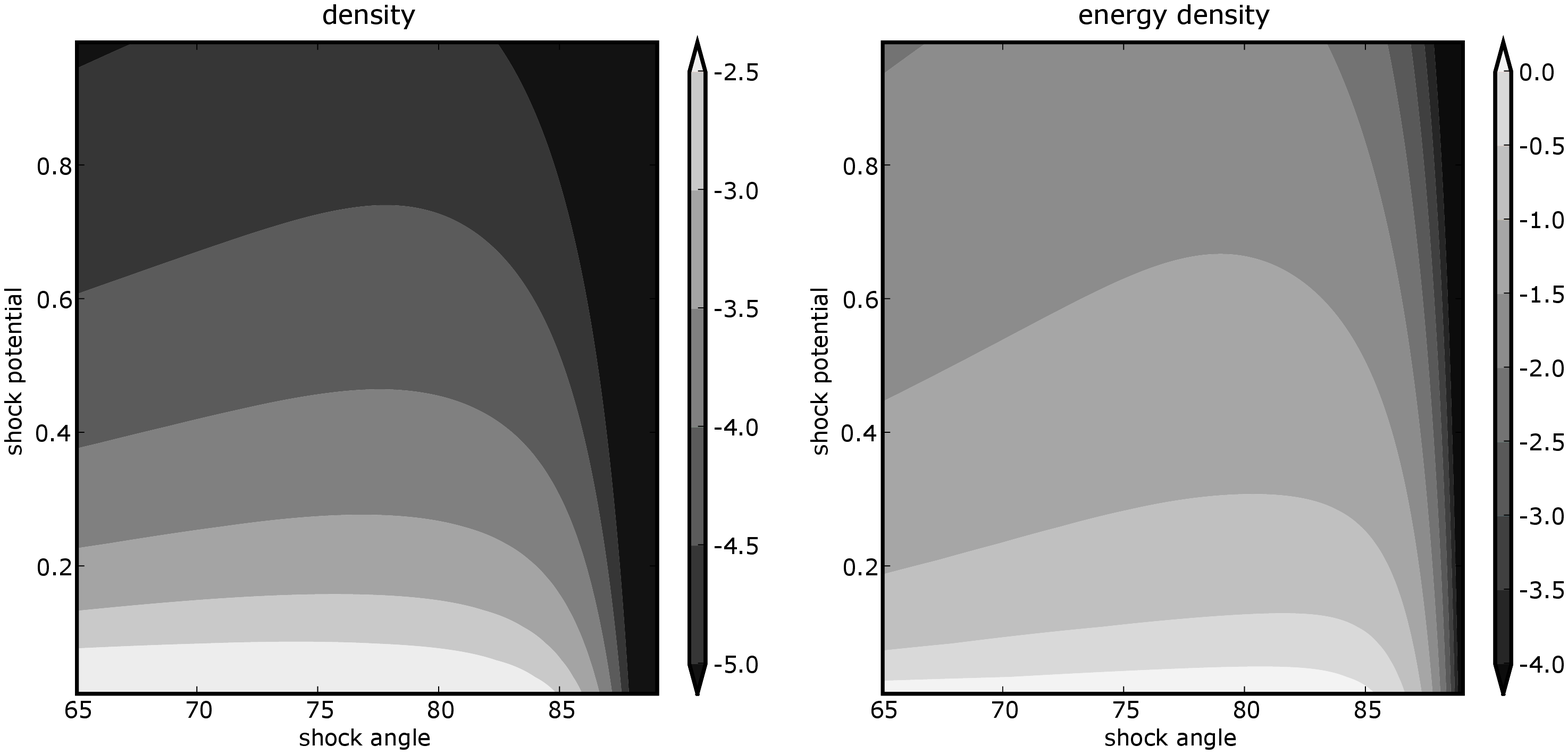} \caption{Density (left)
 and energy density (right) of reflected electrons. The contour levels
 are equally spaced on a logarithmic scale. The vertical axis represents
 the shock potential normalized to the ion bulk energy in the upstream.}
 \label{fig:supernova-model}
\end{figure}

\figref{fig:supernova-model} shows the injection efficiency (left) and
the energy density (right) obtained by the present model. We consider
the shock potential as a free parameter, because it is also difficult to
estimate the precise value of the shock potential at high Mach number
shocks. We expect the potential will be in the range of $\phiht =
0.3-0.5$. Then, we obtain the peak injection efficiency of $\sim 2
\times 10^{-4}$ at $\theta_{Bn} \simeq 80 \degr$. Similarly, the energy
density at the peak shock angle is approximately 10\%.

These results can be directly compared to X-ray and radio observations
of SNRs. \cite{2003ApJ...589..827B} carried out a detailed investigation
of the northeast shell of SN 1006 observed by \Chandra and argued that
the estimated injection efficiency is $\sim 1 \times 10^{-3}$ and the
energy density of nonthermal particles is about $30\%$ of the thermal
energy density. These values are similar to previous observations
\citep[e.g.][]{2001ApJ...558..739A}. The model prediction shows a good
agreement with the observation, although the observed injection
efficiency and the energy density of cosmic ray electrons are slightly
larger than those obtained by the present model. Furthermore, the high
resolution observation reveals that nonthermal emission is confined in
very thin filaments. The spatial scales of the nonthermal filaments are
$\sim 0.04 \pc$ and $\sim 0.2 \pc$ in the upstream and downstream,
respectively. Similar results are also reported by
\cite{2003ApJ...586.1162L}. If the standard DSA theory is employed, the
observed scale length of the nonthermal filaments imposes the constraint
that the shock angle should be sufficiently large ($\theta_{Bn} \gtrsim
80$), unless we assume the magnetic field of $20-85 \mug$, which is
larger than the usual interstellar value of a few $\mug$
\citep[e.g.][]{2003ApJ...589..827B,2004A&A...416..595Y}. The peak shock
angle predicted by the model is around $80\degr$, again we see a good
agreement with the observation.

We should note that the peak shock angle depends on the choice of both
$\Pesc$ and $\Emax$. If $\Emax$ is sufficiently large compared to the
shock potential, the peak shock angle is simply determined by
$\Pesc$. Although we attribute the escape probability to the shock front
nonstationarity in the present analysis, there remain some other
possibilities such as pitch angle scattering due to the interaction with
the whistler waves in the shock transition region. The discussion of
such nonadiabatic behaviors of energetic particles requires more
detailed understandings of the structure and the wave activities in the
shock transition region.

\section{DISCUSSION}
We have studied rapid electron energization mechanism within the
transition region of high Mach number quasi-perpendicular shocks. We
found that highly energetic electrons are generated through successive
two different acceleration processes. First, energetic electrons are
produced via SSA at the leading edge of the shock transition region. As
a result, the preaccelerated electrons escaping outside the loss cone
are subjected to SDA and preferentially reflected back to the
upstream. We consider the two-step acceleration mechanism as the
injection to subsequent DSA process. We have constructed a model of the
acceleration mechanism which predicts the injection efficiency and the
energy density of nonthermal particles. The estimated injection
efficiency agrees well with observations of SN 1006. We also found that
the shock angle dependence of the injection efficiency is consistent
with the shock angle constraint inferred from observations.

Although the present model generally agrees well with observations,
there remain some important issues. The most important one is the
acceleration efficiency of SSA, i.e. the maximum energy and the spectral
index. It is easily understood that the spectral index affect the
injection efficiency to a great extent. We have used the power law index
of $3.5$ throughout in this paper. However, it may depend on some
important physical parameters such as $\wpe/\wce$, $\theta_{Bn}$,
$m_i/m_e$ etc. We also observe that the spectral index varies with time
according to the phase of the self-reformation. The maximum energy of
SSA is also important for the electron injection. The maximum energy,
which is required in order to account for observations, depends on the
shock potential. The maximum energy should be larger than the shock
potential to obtain a measurable fraction of reflected electrons. Even
in this case, the maximum energy is important for determining the shock
angle dependence of the injection efficiency, because it introduces the
cut-off shock angle where the injection efficiency rapidly decreases.

We also introduce the escape probability of energetic electrons as
another important factor for determining the shock angle dependence. We
attribute the probability to the self-reformation of the shock
front. This is because the value of the escape probability estimated
from the simulation results is $\Pesc \simeq 4 \wci$, which indicates
that the escape mechanism is, to some extent, related to the shock
self-reformation process. Although the self-reformation process of the
shock front has been extensively studied by many authors
\citep[e.g.][]{1986JGR....91.8805Q,1992PhFlB...4.3533L,2003JGRA.108a.SSH4S},
it is still a controversial topic of collisionless shock
physics. \cite{2005JGRA..11002105S} have recently pointed out that the
dynamics of the shock front can be modified by strong dissipation due to
microinstabilities in the transition region of very high Mach number
shocks.  Similar phenomenon at moderate Mach number shock $M_A \sim 6$
is also reported by \cite{2004AnGeo..22.2345S}, where the modified two
stream instability plays an important role
\citep[e.g.][]{2003JGRA.108lSMP19M}. It is difficult to say whether the
shock self-reformation process survives in real SNR shocks, where strong
dissipation of both electrons and ions is expected. If the
self-reformation is suppressed by strong dissipation, the escape
probability may become smaller. If this is the case, the cut-off shock
angle determined by the escape probability disappears and the maximum
cut-off energy of SSA controls the shock angle dependence.

If the shock self-reformation is suppressed and the maximum energy of
SSA becomes sufficiently large, the reflection of energetic electrons
will take place at the shock angle very close to the threshold between
subluminal and superluminal shocks. The relativistic effect becomes
important for such situations, because both the effective shock velocity
$\Vsh$ and the required energy for the reflection become
relativistic. In this case, since the shock angle is well close to
$90\degr$, the shock potential has little influence on the reflection
process, i.e., the relativistic effect does not affect the mirror
reflection efficiency in the absence of the electrostatic
potential. Thus, the present mechanism should work in principle. It is
important to investigate whether highly relativistic electrons are
generated via SSA or not. Stochastic version of SSA may be important to
understand the issue \citep{2001PhRvL..87y5002M}.

We should also point out that further electron heating and acceleration
in the shock transition region may be possible if multidimensional
effects are considered. It is well known that many plasma
microinstabilities can be excited within the shock transition region
\cite[e.g.][]{1984SSRv...37...63W}. For instance, it is easy to expect
that the whistler mode waves are excited by temperature anisotropy
($T_{\perp} > T_{\parallel}$) of thermal electron because of strong
perpendicular heating and acceleration due to SSA and the ion acoustic
turbulence. We also expect that the reflected electron beam with
temperature anisotropy ($T_{\perp} > T_{\parallel}$) will excite the
whistler waves propagating antiparallel to the beam by cyclotron
resonance as discussed by \cite{1984JGR....89..105T}, although the use
of one-dimensionality assumption in the present simulation inhibits the
excitation of these instabilities. If the whistler wave intensity
becomes sufficiently strong, both the beam and core electron
distributions will become isotropic via strong pitch angle
scattering. The first adiabatic invariant of electron will be violated
and our simple theoretical model using the adiabatic approximation may
be inaccurate. Even in this case, there are no reason why SDA should not
operate, because the physical mechanism is quite simple and does not
require any special conditions. However, the process will be strongly
modified by the turbulence in the shock transition region. The
wave-particle interaction within the shock layer may provide further
heating and acceleration of electrons. We would like to emphasize again
the significant importance of SSA on the energization of electrons. SSA
plays a key role in the turbulent shock structure in the sense that it
provides additional sources of free energy and may lead to further
energization of electrons.

The relation between the electron acceleration efficiency and the
whistler waves is recently studied by \cite{2006GeoRL..3324104O}. They
analyzed a number of the Earth's bow shock crossing events observed by
Geotail. They clearly showed that the power law index of electron energy
spectra measured in the shock transition region is regulated by the
so-called whistler critical Mach number $M_{crit}^w$, which is defined
as the critical point above which the whistler waves cannot propagate
upstream. The spectral indices are distributed $3.5-5.0$ in the
sub-critical regime, while the harder energy spectra with indices of
$3.0-3.5$ are observed in the super-critical regime. In the
super-critical regime, the whistler waves generated by
microinstabilities are accumulated in the shock transition region. Since
SNR shocks are in super-critical regime, we can expect that the
accumulated energy of the whistler waves may contribute to further
electron energization. It is interesting to investigate the relationship
between the whistler waves and the electron acceleration
efficiency. Gyroresonant surfing acceleration proposed by
\cite{2005PhRvL..94c1102K} might be important.

In order to model such nonadiabatic processes, we must know the shock
structure (e.g. shock potential, nonstationarity) and the wave activity
in the transition region of realistic high Mach number shocks in more
detail. Numerical simulation of self-consistent shock structures
including the whistler wave turbulence requires at least two-dimensional
simulation domain which demands very large computational
resources. Another possibility to improve the understandings of high
Mach number shocks is in situ observation of interplanetary shocks in
the inner heliosphere. It is known that interplanetary shocks driven by
Coronal Mass Ejections (CMEs) can be very high Mach number near the Sun
\citep[e.g.][]{1985JGR....90..183S}. Expected Mach number at the Mercury
orbit of $\sim 0.4$ AU becomes $M_A \sim 40$. Observations of such
interplanetary shocks may provide us useful information to understand
the physics of electron acceleration at very high Mach number shocks.

\begin{figure}
 \figurenum{9} \epsscale{1.0} \plotone{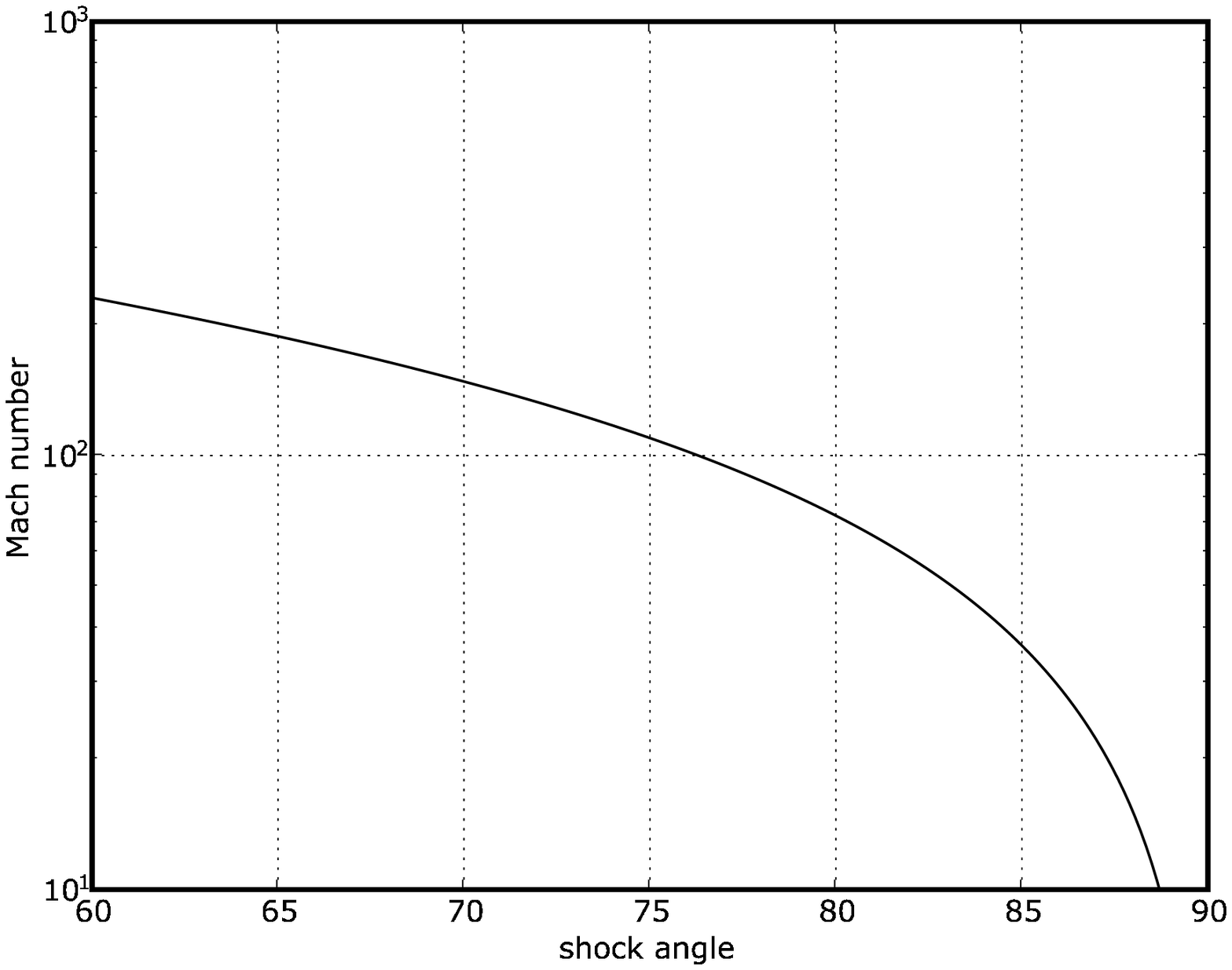} \caption{Critical Mach
 number above which the self-generation of upstream waves becomes
 possible.}  \label{fig:critical-Ma}
\end{figure}

In the present paper, we have restricted ourselves to the discussion of
the electron injection process. The self-generation of upstream waves is
one of the major problems of the electron DSA theory. The cyclotron
resonance condition between the reflected electron beam and a left-hand
polarized weakly damped \Alfven wave ($k v_{A} / \wci \lesssim 1$)
requires,
\begin{eqnarray}
 \frac{v_r}{V_1} \gtrsim \frac{1}{2} \frac{1}{M_A} \frac{m_i}{m_e},
  \label{eq:self-generation-critical}
\end{eqnarray}
where the factor $1/2$ in the right-hand side of the above equation
indicates the difference of the frame between the shock frame and the
upstream frame. If we use typical Mach number of SNR shocks $M_A \sim
100 - 1000$, this condition becomes $v_r/V_1 \gtrsim 1 - 10$. By
combining (\ref{eq:beam-velocity}) and
(\ref{eq:self-generation-critical}), we can estimate the critical Mach
number above which the self-generation of upstream waves becomes
possible. \figref{fig:critical-Ma} shows the shock angle dependence of
the critical Mach number. We confirm that the self-generation of
upstream waves is indeed possible in typical SNR shocks, i.e., our
theory successfully explains the electron injection as well as the
triggering of subsequent DSA process.

Since we also know the reflected electron density, we can, in principle,
estimate the wave intensity, hence, the diffusion coefficient of
energetic electrons using quasi-linear theory. However, we think that it
is rather important to investigate further electron energization in the
shock transition region introduced by multidimensional effect before we
discuss the consequence of the electron injection and subsequent DSA
process.

Another important problem is the back-reaction from nonthermal
particles. It is known that the energetic particles affect the upstream
plasma environment when their energy density becomes comparable to that
of the background plasma \citep[e.g.][]{1981ApJ...248..344D}. We cannot
discuss such a nonlinear evolution in the present model, because we do
not consider the injection process of ions. Obviously, ions have much
larger energy density than that of electrons. Thus, the injection
efficiency of ions is more important to understand the shock structure
in the presence of energetic particles. In order to study the nonlinear
evolution including the interaction between thermal and nonthermal
particles, the injection efficiencies of both electrons and ions will be
of great importance. Understanding of both the injection processes and
the nonlinear shock structure will elucidate the problem of cosmic ray
acceleration at collisionless shock waves.

\acknowledgments This work is supported by ISAS/JAXA and the
Solar-Terrestrial Environment Laboratory, Nagoya University. T.A. is
supported by JSPS Research Fellowship for Young Scientists.

\appendix
\section{Calculation of Reflection Time}
The parallel equation of motion in the HTF under the action of the
magnetic mirror force and the electrostatic potential can be written as
\begin{eqnarray}
 m_e \frac{d v_{\parallel}}{d t} = - \left( \mu \frac{\partial
  B}{\partial x} + e \frac{\partial \phi}{\partial x} \right) \cos
  \theta_{Bn},
\end{eqnarray}
where $x$ represents the particle position and $\mu = m_e v_{\perp}^2/2
B$ is the first adiabatic invariant. By assuming the spatial gradients
of both the magnetic field strength and the electrostatic potential are
constant, we can integrate the equation,
\begin{eqnarray}
 x(t) = v_{\parallel}(0) t - \frac{\cos \theta_{Bn}}{2 m_e} \frac{\mu
  \Delta B - e \Delta \phi}{\Delta x} t^2.
\end{eqnarray}
In this calculation, we have neglected the curvature of the magnetic
field line in the shock transition region for simplicity (i.e. $\cos
\theta_{Bn} = const.$). The initial parallel velocity is
$v_{\parallel}(0) \simeq \Vsh$ and the shock width is approximately
given by $\Delta x \simeq V_1/\wci$. Then, we obtain the characteristic
time for the reflection process $\Tref$ as
\begin{eqnarray}
 \wci \Tref \simeq \frac{2 m_e \Vsh^2}{\mu \Delta B - e \Delta \phi}.
\end{eqnarray}
If we use the definition of the characteristic particle energy
perpendicular to the magnetic field before the reflection $\Eref = m_e
v_{\perp}^2 / 2$,
\begin{eqnarray}
 \mu \Delta B &=& \Eref \frac{B_{\rmax} - B_{\foot}}{B_{\foot}}
  \nonumber \\ &=& \frac{\Eref}{\tan^2 \theta_c}.
\end{eqnarray}
Thus, we can rewrite the reflection time
\begin{eqnarray}
 \wci \Tref \simeq \frac{2 m_e \Vsh^2}{\Eref / \tan^2 \theta_c - e
 \Delta \phi}.
\end{eqnarray}
It is easy to understand that $\Tref$ is proportional to $1/\cos^2
\theta_{Bn}$ for the fixed particle energy. The reflection time becomes
shorter with increasing the particle energy.


\end{document}